\documentclass[12pt]{iopart}

\usepackage{iopams}
\usepackage[square,sort&compress,numbers]{natbib}
\usepackage{graphicx}
\usepackage{epsfig}
\usepackage{epstopdf}
\usepackage{color}
\usepackage{ulem} 
\begin{document}

\title{Artificial intelligence meets minority game: toward optimal resource allocation}

\author{Si-Ping Zhang}
\address{Institute of Computational Physics and Complex Systems, Lanzhou University, Lanzhou 730000, China}

\author{Jia-Qi Dong}
\address{Institute of Computational Physics and Complex Systems, Lanzhou University, Lanzhou 730000, China}

\author{Li Liu}
\address{School of Software Engineering, Chongqing University, Chongqing 400044, PR China}

\author{Zi-Gang Huang}
\address{School of Life Science and Technology, Xi'an Jiao Tong University, Xi'an 710049 China}
\ead{huangzg@xjtu.edu.cn}

\author{Liang Huang}
\address{Institute of Computational Physics and Complex Systems, Lanzhou University, Lanzhou 730000, China}

\author{Ying-Cheng Lai}
\address{School of Electrical, Computer and Energy Engineering, Department of Physics, Arizona State University, Tempe, AZ 85287, USA}

\date{\today}

\clearpage

\begin{abstract}

Complex resource allocation systems provide the fundamental support for the
normal functioning and well being of the modern society. Computationally and
mathematically, such systems can be modeled as minority games. A ubiquitous
dynamical phenomenon is the spontaneous emergence of herding, where a vast
majority of the users concentrate on a small number of resources. From an
operational point of view, herding is of grave concern as the few overused
resources can be depleted quickly, directing users to the next few resources
and causing them to fail, and so on, and eventually leading to a catastrophic
collapse of the whole system in short time. To devise strategies to prevent
herding from occurring is thus of interest. Previous works focused on
control strategies that rely on external interventions, such as pinning
control where a fraction of users are forced to choose a certain action.
Is it possible to eliminate herding without any external control? The
main point of this paper is to provide an affirmative answer through
exploiting artificial intelligence (AI). In particular,
we demonstrate that, when agents are empowered with reinforced
learning (e.g., the popular Q-learning in AI) in that they get familiar with
the unknown game environment gradually and attempt to deliver the optimal
actions to maximize the payoff, herding can effectively be eliminated. Furthermore, computations reveal the striking
phenomenon that, regardless of the initial state, the system evolves
persistently and relentlessly toward the optimal state in which all resources
are used efficiently. However, the evolution process is not without
interruptions: there are large fluctuations that occur but only intermittently
in time. The statistical distribution of the time between two successive
fluctuating events is found to depend on the parity of the evolution, i.e.,
whether the number of time steps in between is odd or even. We develop a
physical analysis and derive mean-field equations to gain an understanding of
these phenomena. As minority game dynamics and the phenomenon of herding are
common in social, economic, and political systems, and since AI is becoming
increasingly widespread, we expect our AI empowered minority game system to
have broad applications.

\end{abstract}

\vspace{2pc}

\maketitle

\section{Introduction} \label{sec:intro}

The tremendous development of information technology has made it possible for
artificial intelligence (AI) to penetrate into every aspect of the human
society. One of the fundamental traits of AI is decision making - individuals,
organizations, and governmental agencies tend to rely more and more
on AI to make all kinds of decisions based on vast available information in
an ever increasingly complex environment. At the present, whether a strong
reliance on AI is beneficial or destructive to the mankind is an issue of
active debate that attracts a great deal of attention from all the professions.
In the vast field of AI related research, a fundamental issue is how AI affects or harnesses the behaviors of complex dynamical systems. In this paper, we
address this issue by focusing on complex resource allocation systems that
incorporate AI in decision making at the individual agent level, and
demonstrate that AI can be quite advantageous for complex systems to reach
their optimal states.

Resource allocation systems are ubiquitous and provide fundamental support for
the modern economy and society, which are typically complex systems consisting
of a large number of interacting elements. Examples include ecosystems of
different sizes, various transportation systems (e.g., the Internet, urban
traffic systems, rail and flight networks), public service providers (e.g.,
marts, hospitals, and schools), as well as social and economic organizations
(e.g., banks and financial markets). In a resource allocation system, a large
number of components/agents compete for limited public resources in order to
maximize payoff. The interactions among the agents can lead to extremely
complex dynamical behaviors with negative impacts on the whole system, among
which irrational herding is of great concern as it can cause certain resources
to be overcrowded but leave others unused and has the potential to lead to
a catastrophic collapse of the whole system in relatively short time.
A general paradigm to investigate the collective dynamics of
resource allocation systems is complex adaptive systems
theory~\cite{Kauffman:book,Levin:1998,AAP:book}. At the microscopic level,
multi-agent models such as the minority game model~\cite{CZ:1997} and
interaction models based upon the traditional game theory~\cite{NPS:2000,
RCS:2009,PD:2012} have been proposed to account for the interactions among
the individual agents.

Minority game is a paradigmatic model for resource allocation in population,
which was introduced in 1997~\cite{CZ:1997} for quantitatively studying the
classic El Farol bar-attendance problem first conceived by Arthur in
1994~\cite{Arthur:1994}. In the past two decades, minority game and its
variants were extensively studied~\cite{CM:1999,SMR:1999,PBC:2000,CMM:2008,
Moro:2004,CMZ:2005,YZ:2008,ZWZYL:2005,EZ:2000,TCHJ:2005,JHH:1999,HJJH:2001,
Marsili:2001,BMFM:2008,XWHZ:2005,ZZZH:2005,ATBK:2004,LCHJ:2004,Slanina:2001,
KSB:2000,MMM:2004,BMM:2007,AK:2002,ZHDHL:2013,ZHWSL:2016,HZDHL:2012,
DHHL:2014}, where a central goal was to uncover the
dynamical mechanisms responsible for the emergence of various collective
behaviors. In the original minority game model, an individual's scheme for
state updating (or decision making) is essentially a trial-and-error learning
process based on the global historical winning information~\cite{CZ:1997}. In
other models, learning mechanisms based local information from neighbors were
proposed~\cite{PBC:2000,ZWZYL:2005,EZ:2000,CMM:2008,ATBK:2004,KSB:2000,
AK:2002,ZHDHL:2013,ZHWSL:2016,HZDHL:2012,DHHL:2014}. The issue of controlling
and optimizing complex resource allocation systems was also
investigated~\cite{ZHDHL:2013}, e.g., utilizing pinning control to harness
the herding behavior, where it was demonstrated that a small number of control
points in the network can suppress or even eliminate herding.
A theoretical framework for analyzing and predicting the efficiency of
pinning control was developed~\cite{ZHDHL:2013}, revealing that the connecting
topology among the agents can play a significant role in the control outcome.
Typically, control requires external interventions. A question is whether
herding can be suppressed or even eliminated without any external control.

In this paper, we address the question of how AI can be exploited to
harness undesired dynamical behaviors to greatly benefit the operation
of the underlying complex system. More generally, we aim to study how
AI affects the collective dynamics in complex systems. For this purpose,
we introduce a minority game model incorporating AI at the individual agent
level, where the agents participating in the game are ``intelligent'' in
the sense that they are capable of reinforced learning~\cite{AA:book}, a
powerful learning algorithm in AI. Empowered with reinforced learning, an
agent is capable of executing an efficient learning path toward a pre-defined
goal through a trial-and-error process in an unfamiliar game environment.
Our model is constructed based on the interplay of a learning agent and the
environment in terms of the states, actions, rewards, and decision making.
In reinforced learning, the concepts of value and value functions are key to
intelligent exploration, and there have been a number of reinforced learning
algorithms, such as dynamic programming~\cite{AA:book,Bellman:1957}, Monte
Carlo method~\cite{AA:book,Bellman:1957}, temporal differences~\cite{AA:book,
Sutton:1998}, Q-Learning~\cite{AA:book,Watkins:1989,WD:1992},
Sarsa~\cite{AA:book}, and Dyna~\cite{AA:book}, etc. To be illustrative, we
focus on Q-learning, which was demonstrated previously to perform well for a
small number of individuals in their interaction with an unknown
environment~\cite{AA:2012,PA:2003,SC:2003,AA:2012,AA:2013}. However, here
we consider minority game systems with a large number of ``intelligent''
players, where Q-learning is adopted for state updating in a stochastic
dynamical environment. The question is whether the multi-agent AI minority
game system can self-organize itself to generate optimal collective behaviors.
Our main result is an affirmative answer to this question. Particularly,
we find that the population of AI-empowered agents can approach the optimal
state of resource utilization through self-organization regardless of the
initial state, effectively eliminating herding. However, the process of
evolution toward the optimal state is typically disturbed by intermittent,
large fluctuations (oscillations) that can be regarded as failure events.
There can be two distinct types of statistical distributions of the
``laminar'' time intervals in which no failure occurs, depending on their
parity, i.e., whether the number of time steps between two consecutive
failures is odd or even. We develop a physical analysis and use the mean-field
approximation to understand these phenomena. Our results indicate that
Q-learning is generally powerful in optimally allocating resources to agents
in a complex interacting environment.

\section{Model} \label{sec:model}

Our minority game model with agents empowered by Q-learning can be described,
as follows. The system has $N$ agents competing for two resources denoted by
$r=+1$ and $-1$, and each agent chooses one resource during each round of the
game. The resources have a finite capacity $C_r$, i.e., the maximum number of
agents that each resource can accommodate. For simplicity, we set $C_r=N/2$.
Let $A(t)$ denote the number of agents selecting the resource $r =+1$ at time
step $t$. For $A(t)\le C_r$, agents choosing the resource $+1$ belong to the
minority group, and win the game in this round. Conversely, for $A(t)> C_r$,
the resource $+1$ is overcrowded, so the corresponding agents fail in this
round.

The Q-learning adaptation mechanism~\cite{WD:1992} is incorporated into the
model by assuming that the states of the agents are parameterized through
$Q$ functions that characterize the relative utility of a particular action.
The $Q$ functions are updated during the course of the agents' interaction
with the environment. Actions that lead to a higher reward are reinforced.
To be concrete, in our model, agents are assumed to have four available
actions, and we let $Q(s,a)$ be the $Q$ value of the corresponding action
at time $t$, where $s$ and $a$ denote the current state of agent and the
action that the agent may take, respectively. A $Q$ function can then be
expressed in the following form:
$$\mathbf{Q}={ \left[ \begin{array}{ccc}
Q(+1,+1) & Q(+1,-1)\\
Q(-1,+1) & Q(-1,-1)
\end{array}
\right ]}.$$
For an agent in state $s$, after selecting a given action $a$, the
corresponding $Q$ value is updated according to the following rule:
\begin{equation} \label{eq:Q_rule}
Q_t(s,a)=Q_{t-1}(s,a)+\alpha[R+\gamma Q_{t-1}^{\mathrm{max}}(s',a')
-Q_{t-1}(s,a)],
\end{equation}
where $\alpha\in(0,1]$ is the learning rate and $R$ is the reward from the
corresponding action. The parameter $\gamma\in[0,1)$ is the discount factor
that determines the importance of future reward. Agents with $\gamma=0$ are
``short sighted'' in that they consider only the current reward, while those
with larger values of $\gamma$ care about reward in the long run. The quantity
$Q_{t-1}^{\mathrm{max}}(s',a')$ is the maximum element in the row of the $s'$
state, which is the outcome of the action $a$ based on $s$.
Equation~(\ref{eq:Q_rule}) indicates that the matrix $\mathbf{Q}$ contains
information about the accumulative experience from history, where the reward
$R$ (for action $a$ from state $s$) and the expected best value
$Q_{t-1}^{\mathrm{max}}(s',a')$ based on $s'$ both contribute to the updated
value $Q_{t}(s,a)$ with the weight $\alpha$, and the previous value
$Q_{t-1}(s,a)$ is also accumulated into $Q_{t}(s,a)$ with the weight
$1-\alpha$.

While agents select the action mostly through reinforced learning, certain
randomness can be expected in decision making. We thus assume that a random
action occurs with a small probability $\epsilon$, and agents select the
action with a large value of $Q(s,a)$ with probability $1-\epsilon$.
For a given setting of parameters $\alpha$ and $\gamma$, the Q-learning
algorithm is carried out, as follows. Firstly, we initialize the matrix
$\mathbf{Q}$ to zero to mimic the situation where the agents are unaware of
the game environment, and initialize the state $s$ of each agent randomly to
$+1$ or $-1$. Next, for each round of the game, each agent chooses an action
$a$ with a larger value of $Q_{t}(s,a)$ in the row of its current state $s$
with probability $1-\epsilon$, or chooses an action $a$ randomly with
probability $\epsilon$. The $Q(s,a)$ value of the selected action is then
updated according to Eq.~(\ref{eq:Q_rule}). The action that leads to the
state $s'$ identical to the current winning (minority) state has $R=1$,
and the action leading to the failed (majority) state has $R=0$. Finally,
we take the selected action $a$ to update the state from $s$ to $s'$.

Distinct from the standard supervised learning~\cite{DW:2016}, agents
adopting reinforced learning aim to understand the environment and
maximize their rewards gradually through a trial-and-error process. The
coupling or interaction among the agents is established through competing for
limited resources. Our AI based minority game model also differs from the
previously studied game systems~\cite{ZHDHL:2013} in that our model takes
into account agents' complicated memory and decision making process. For our
system, a key question is whether the resulting collective behaviors from
reinforced learning may lead to high efficiency or optimal resource
allocation in the sense that the number of agents that a resource
accommodates is close to its capacity.

\section{Self-organization and competition} \label{sec:RL}

\begin{figure}[h!]
\centering
\includegraphics[width=\linewidth]{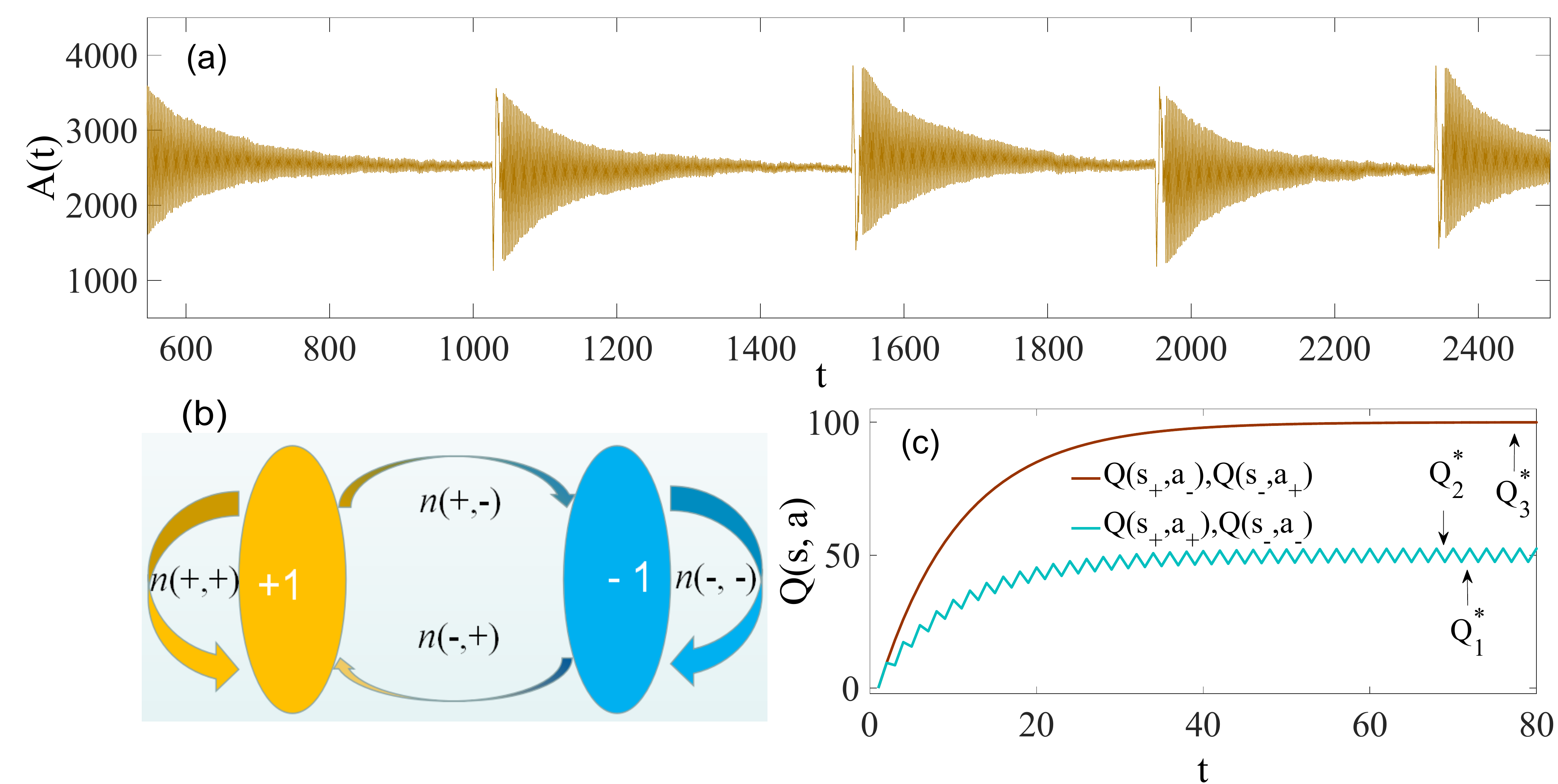}
\caption{ \textbf{Typical temporal evolutionary behavior of the proposed
AI minority game system with reinforced learning empowered agents.}
(a) Time series of the attendance $A(t)$ of resource $+1$. Interactions among
the agents make the system self-organize into a special temporal pattern with
two main features: the convergence of regular oscillations towards the
optimal value $C_r=N/2$, and intermittent bursts of failures in utilizing
resources. (b) A schematic sketch of the state transitions of agents during
the dynamical process. There are self-satisfied agents in a fixed state and
speculative agents that continuously switch state between $+1$ and $-1$.
(c) Time series of $Q(s,a)$ as the numerical solutions of
Eqs.~(\ref{eq:Qsa+}-\ref{eq:Qsas}). The parameters are: learning rate
$\alpha=0.9$, discount factor $\gamma=0.9$, exploration rate $\epsilon=0.02$,
and system size $N=5001$.}
\label{fig:sketch}
\end{figure}

In the traditional minority game, the dynamical rules stipulate that
competition and learning among agents can lead to the detrimental herding
behavior, to which game systems composed of less diversified agents are
particularly susceptible~\cite{ZHWSL:2016,ZHDHL:2013,HZDHL:2012,DHHL:2014}.
In our AI minority game system of agents empowered with reinforced learning,
herding is dramatically suppressed. To give a concrete example, we set the
parameters for Q learning as: learning rate $\alpha=0.9$, discount factor
$\gamma=0.9$, and exploration rate $\epsilon=0.02$. Figure~\ref{fig:sketch}(a)
shows the temporal evolution of the number $A(t)$ of agents choosing resource
$+1$. The main features of the time series are the continuous oscillations
of $A(t)$ about the capacity $C_r$ of resources, convergence of the oscillation
amplitude, and bursts of $A(t)$ that occur intermittently. As the oscillations
converge to the optimal state, the two resources $r=+1$ and $r=-1$ play as the
minority resource alternatively. The remarkable feature is that the agent
population tends to self-organize into a non-equilibrium state with certain
temporal pattern in order to reach the highly efficient, optimal state, but
the process is interrupted by large bursts (failures or fluctuations).

\subsection{Convergence of oscillations}

\paragraph*{Emergence of two types of agents.}
From numerical simulations of the AI minority game system, we find that,
as the system self-organizes itself into patterns of regular oscillations,
agents with two types of behaviors emerge. The first type is those agents
who are ``self-satisfied'' in the sense that they remain in either the
$s=+1$ state or the $s=-1$ state. Those agents win and lose the game
alternatively as the system develops regular oscillations. The population
sizes of the self-satisfied agents are denoted as $n(+1,+1)$ and $n(-1,-1)$,
respectively. The second type of agents are the ``speculative'' agents,
or speculators, who switch state at each time step between $s=+1$ and
$s=-1$. These agents always win the game when the system exhibits regular
oscillations. We denote the population sizes of the speculative agents as
$n(+,-)$ and $n(-,+)$, which correspond to the two possibilities of
switching: from $s=+1$ to $s=-1$ and vice versa, respectively.

Figure~\ref{fig:sketch}(b) shows the state transition paths induced by the
self-satisfied agents and the speculative agents. The oscillations of $A(t)$
associated with the convergent process can be attributed to the state
transition of the speculative agents between the states $+1$ and $-1$. This
agrees with the intuition that, e.g., the investing behavior of speculators
in a financial market is always associated with high risks and large
oscillations. Due to the decrease in the population of the speculative agents,
the oscillation amplitude in any time interval between two successive failure
events tends to decay with time.

\paragraph*{Stable state of Q table.}
The oscillations of $A(t)$ mean that $r=+1$ and $-1$ act as the minority
resource alternatively. For the self-satisfied agents, according to the
Q-learning algorithm, the update of the element $Q(s_+,a_+)$ can be
expressed as,
\begin{equation} \label{eq:Qsa+}
\left\{
\begin{array}{l}
Q_{t+1}(s_+,a_+) = Q_t(s_+,a_+)+\alpha[R+\gamma Q_t(s_+,a_+)-Q_t(s_+,a_+)]\\
Q_{t+2}(s_+,a_+) = Q_{t+1}(s_+,a_+)+\alpha[\gamma Q_{t+1}(s_+,a_+)
-Q_{t+1}(s_+,a_+)]\\
\end{array},
\right.
\end{equation}
where $Q_t^{\mathrm{max}}(s',a')=Q_t(s_+,a_+)$ due to the inequality
$Q(s_+,a_+)>Q(s_+,a_-)$. The update of the element $Q(s_-,a_-)$ is described by
\begin{equation} \label{eq:Qsa-}
\left\{
\begin{array}{l}
Q_{t+1}(s_-,a_-) = Q_t(s_-,a_-)+\alpha[R+\gamma Q_t(s_-,a_-)-Q_t(s_-,a_-)]\\
Q_{t+2}(s_-,a_-) = Q_{t+1}(s_-,a_-)+\alpha[\gamma Q_{t+1}(s_-,a_-)
-Q_{t+1}(s_-,a_-)]\\
\end{array},
\right.
\end{equation}
where $Q_t^{\mathrm{max}}(s',a')=Q_t(s_-,a_-)$ as a result of the inequality
$Q(s_-,a_-)>Q(s_-,a_+)$.

For the speculative agents, the updating equations of elements $Q(s_+,a_-)$ and
$Q(s_-,a_+)$ are
\begin{equation} \label{eq:Qsas}
\left\{
\begin{array}{l}
Q_{t+1}(s_+,a_-) = Q_t(s_+,a_-)+\alpha[R+\gamma Q_t(s_-,a_+)-Q_t(s_+,a_-)]\\
Q_{t+2}(s_-,a_+) = Q_{t+1}(s_-,a_+)+\alpha[R+\gamma Q_{t+1}(s_+,a_-)-Q_{t+1}(s_-,a_+)],\\
\end{array}
\right.
\end{equation}
where $Q_t^{\mathrm{max}}(s',a')=Q_t(s_-,a_+)$ or $Q_t(s_+,a_-)$, due to the
inequalities $Q(s_+,a_+)<Q(s_+,a_-)$, and $Q(s_-,a_-)<Q(s_-,a_+)$.

Figure~\ref{fig:sketch}(c) shows numerically obtained time series of the
elements of the matrix $\mathbf{Q}$ from Eqs.~(\ref{eq:Qsa+}-\ref{eq:Qsas}).
For the self-satisfied agents, the values of $Q(s_+,a_+)$ and $Q(s_-,a_-)$
increase initially, followed by an oscillating solution between the two
values $Q_1^*$ and $Q_2^*$, where
\begin{eqnarray}
\nonumber
Q_1^* & = & \frac{[1+\alpha(\gamma-1)]\alpha R}{1-[1+\alpha (\gamma-1)]^2}
\\ \nonumber
\mbox{and} \ \ Q_2^* & = & \frac{\alpha R}{1-[1+\alpha (\gamma-1)]^2}
\end{eqnarray}
are obtained from
Eqs.~(\ref{eq:Qsa+}) and~(\ref{eq:Qsa-}). For the speculative agents, both
$Q(s_+,a_-)$ and $Q(s_-,a_+)$ reach a single stable solution
$Q_3^*=R/(1-\gamma)$, which can be obtained by solving
Eq.~(\ref{eq:Qsas}). The three relevant values have the relationship
$Q_1^*<Q_2^*<Q_3^*$.

The emergence of the two types of agents can be understood from the following
heuristic analysis. In the dynamical process, a speculative agent emerges
when the element associated with an agent satisfies the inequalities
$Q(s_+,a_-)>Q(s_+,a_+)$ and $Q(s_-,a_+)>Q(s_-,a_-)$ simultaneously. Initially,
the agents attend both resources $+1$ and $-1$, with one group winning but
the other losing. Only the group that always wins the game can reinforce
themselves through further increment in $Q(s_+,a_-)$ and $Q(s_-,a_+)$. The
stable group of speculative agents leads to regular oscillations of $A(t)$,
because they switch states together between $+1$ and $-1$. An agent becomes
self-satisfied when it is in the $+1$ state and the inequality
$Q(s_+,a_+)>Q(s_+,a_-)$ holds, or in the $-1$ state and
$Q(s_-,a_-)>Q(s_-,a_+)$ holds. The self-satisfied state can be strengthened
following the evolution governed by Eqs.~(\ref{eq:Qsa+}) and (\ref{eq:Qsa-}),
with $Q(s_+,a_+)$ or $Q(s_-,a_-)$ reaching the oscillating state between
$Q_1^*$ and $Q_2^*$, as shown in Fig.~\ref{fig:sketch}(c). We see that
the condition for an agent to become speculative is more strict than to be
self-satisfied. Moreover, a speculative agent has certain probability to
become self-satisfied, as determined by the value of the exploration rate
$\epsilon$. As a result, the population of the speculative agents tends to
shrink, leading to a decrease in the oscillation amplitude $|n(+,-)-n(-,+)|$
and convergence of $A(t)$ closer to the optimal state $N/2$.

\begin{figure}[h!]
\centering
\includegraphics[width=\linewidth]{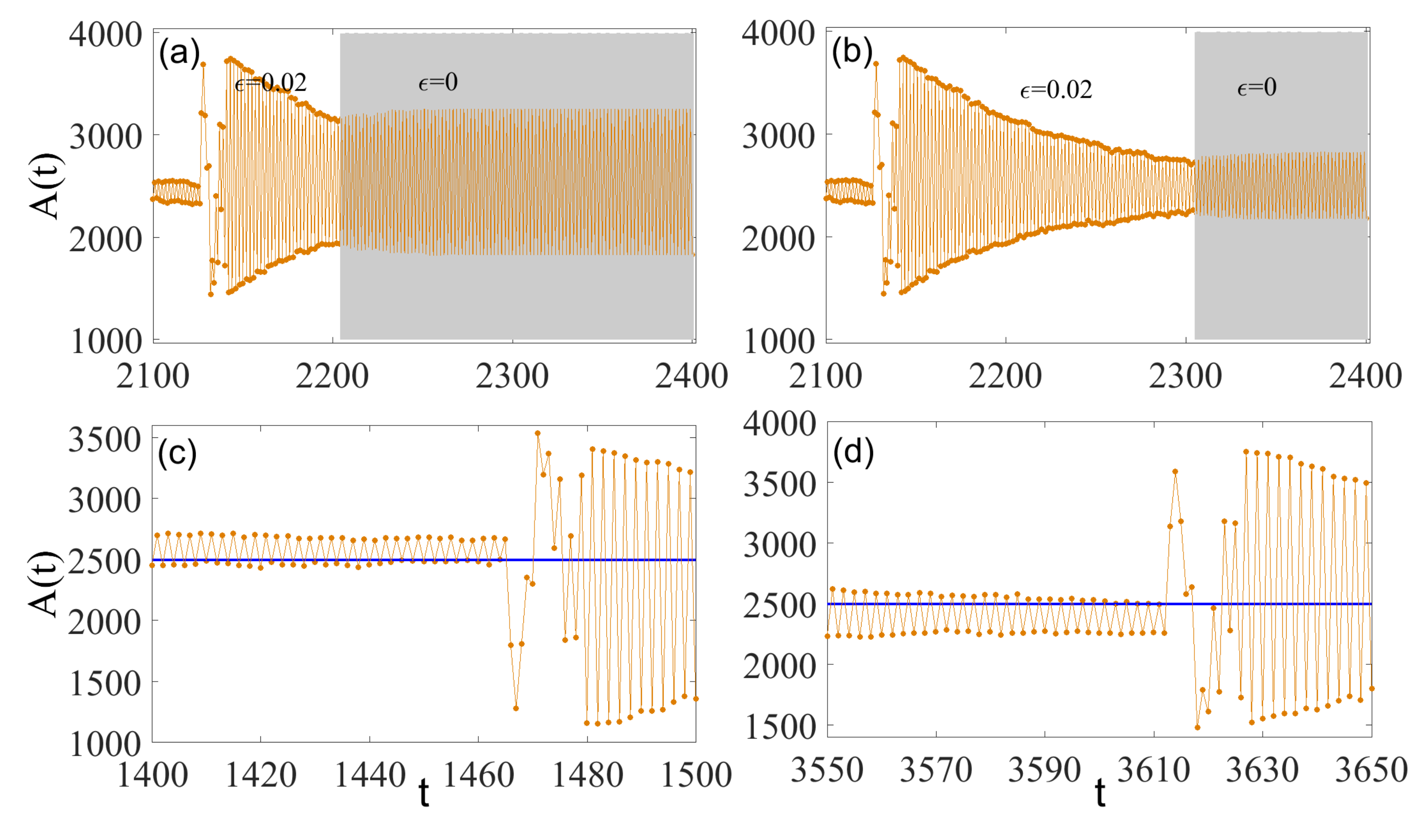}
\caption{ \textbf{Convergence of regular oscillations and bursts of failure.}
(a,b) Convergence of the regular oscillation pattern depends on the
exploration behavior of the agents as characterized by the rate $\epsilon$.
For $\epsilon=0$ (gray region), the oscillation amplitude does not converge.
(c,d) Detailed processes for the bursts of failure. If the regular
oscillations do not cross the line $C_r=N/2$ but behave either as
(c) $A(t)>C_r$ and $A(t+1)>C_r$ or as (d) $A(t)<C_r$ and $A(t+1)<C_r$, the
regular oscillations stop and a systematic failure burst emerges. The blue
line specifies $A(t)=C_r$. The parameters are the same as in
Fig.~\ref{fig:sketch}.}
\label{fig:jump}
\end{figure}

For the special case of $\epsilon = 0$ [the gray regions in
Figs.~\ref{fig:jump}(a) and \ref{fig:jump}(b)], agents take action entirely
based on historical experience $\mathbf{Q}$. In this case, the numbers of
the self-satisfied and speculative agents become constant, and $A(t)$ no
longer converges to that associated with the optimal state. It is thus
apparent that exploration plays a crucial role in the convergence of the
system dynamics toward the state in which the resources are optimally
utilized.

\subsection{Intermittent failures in the AI minority game system}

The intermittent bursts of failure events in the whole system take place
during the convergent process to the optimal state. An understanding of the
mechanism of the failures can provide insights into the articulation of
strategies to make the system more robust and resilient.

The criterion to determine if an agent selecting $+1$ wins the minority game
is $A(t)<C_r=N/2$. If the event $A(t)<C_r$ [or $A(t)>C_r$] occurs twice in
row, the oscillation pattern will be broken. Since the agents are empowered
with reinforced learning, two consecutive winnings of either resource $-1$
or resource $+1$ represent an unexpected event, and this would lead to
cumulative errors in the $Q$ table, triggering a burst of error
in decision making and, consequently, leading to failures in utilizing the
resources. To see this in a more concrete way, we note that a self-satisfied
agent wins and fails alternatively following a regular oscillation pattern.
If the agent fails twice in row, its confidence in preserving the current
state is reduced. As a result, the event $Q(s_+,a_+)<Q(s_+,a_-)$ or
$Q(s_-,a_-)<Q(s_-,a_+)$ would occur with a high probability, leading to a
decrease in the populations $n(+,+)$ and $n(-,-)$ of the self-satisfied
agents. The populations of the speculative agents, $n(+,-)$ and $n(-,+)$,
are increased accordingly. These events collectively generate a bursting
disturbance to the regular oscillation pattern of $A(t)$, terminating the
system's convergence toward the optimal state, as shown in
Figs.~\ref{fig:jump}(a) and \ref{fig:jump}(b).

In general, the stability of the regular oscillations depends on two factors:
the equilibrium position determined by the self-satisfied agents, and the
random fluctuations introduced by agents' exploration behavior. For the
first factor, the equilibrium position is given by
$A_{0}=n(+,+)+[n(-,+)-n(+,-)]/2$, which deviates from $C_r$ due to the
asymmetric distribution of the self-satisfied agents in the two distinct
resources. Figures~\ref{fig:jump}(a) and \ref{fig:jump}(b) show two examples
with the equilibrium position $A_0$ larger or smaller than $C_r$ (the blue
solid line), respectively. We see that the converging process is terminated
when either the upper or the lower envelope reaches $C_r$, i.e., when two
consecutive steps of $A(t)$ stay on the same side of $C_r$ in replacement of an
oscillation about $C_r$. In the thermodynamic limit, for an infinitely large
system with self-satisfied agents symmetrically distributed between $+1$
and $-1$ (so that the equilibrium position $A_0$ is at $C_r$), the oscillation
would persist indefinitely and $A(t)$ approaches $C_r$ asymptotically.

The second factor of random fluctuations in agents' exploratory behavior is
caused by the finite system size, which affects the oscillation stability.
As the populations [$n(+,-)$ and $n(-,+)$] of the speculative agents decrease
during the converging process, the amplitude of oscillation, $|n(+,-)-n(-,+)|$,
becomes comparable to $\sqrt{\epsilon N}$, the level of random fluctuations
in the system. The occurrence of two consecutive steps of $A(t)>C_r$ (or
$A(t)<C_r$) as a result of the fluctuations will break the regular
oscillation pattern. In the thermodynamical limit, the effects of the random
fluctuations are negligible.

\subsection{Time intervals between failure bursts}

\begin{figure}[h!]
\centering
\includegraphics[width=\linewidth]{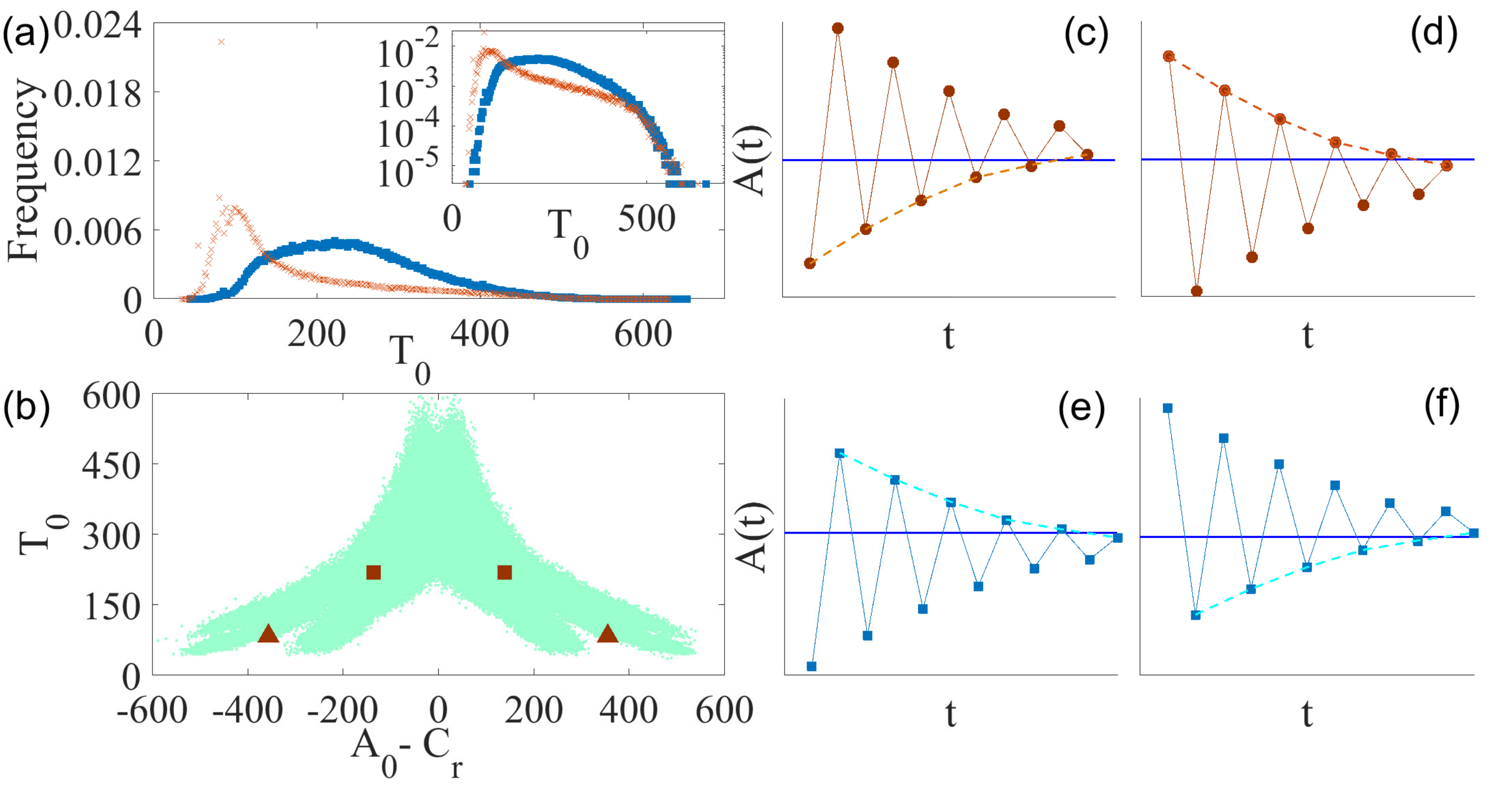}
\caption{ \textbf{Statistical distributions of the time interval $T_0$ between
two successive bursts of failure.} (a) The distributions obtained from one
realization of the system dynamics, where those of the odd $T_0$ values (red
crosses) and even $T_0$ values (blue squares) are remarkably distinct. (b)
$T_0$ versus the deviation $A_0-C_r$ of the equilibrium position from the
resource capacity. The solid squares (triangles) denote the most probable
value of the set of even (odd) $T_0$ values. The parameters are $\alpha=0.9$,
$\gamma=0.9$, and $N=5001$. (c-f) Schematic illustration of four cases
associated with the regular oscillations of $A(t)$, where cases (c,d) lead to
odd intervals $T_0$ while cases (e,f) lead to the even values of $T_0$. The
dashed curves represent the envelopes that cross the capacity value $C_r$
(solid blue lines), which triggers a failure burst.}
\label{fig:stable}
\end{figure}

The dynamical evolution of the system can be described as random failure
bursts superimposed on regular oscillations with decreasing amplitude. The
intermittent failures can be characterized by the statistical distribution
of the time interval $T_0$ between two successive bursting events.
Figure~\ref{fig:stable}(a) shows a representative histogram of $T_0$ obtained
from a single statistical realization of the system dynamics (the inset's
showing the same data but on a semi-logarithmic scale). A
remarkable feature is that the distributions of the odd (red crosses) and even
values of $T_0$ (blue squares) are characteristically distinct. In particular,
the odd values of $T_0$ emerge with a smaller probability and the corresponding
distribution has a smaller most probable value as compared with that for the
even values of $T_0$. A possible explanation lies in the existence of two
intrinsically distinct processes.

Our computation and analysis indicate that the regular oscillation processes
can be classified into two categories, as shown in Figs.~\ref{fig:stable}(c-f),
leading to insights into the mechanism for the two distinct types of
statistical distributions in $T_0$. In Fig.~\ref{fig:stable}(c), $A(t)$ starts
from a value below $C_r=N/2$ and terminates at a value above $C_r$, due to the
two consecutive values above $C_r$ as the lower envelope of $A(t)$ crosses
$C_r$. Similarly, in Fig.~\ref{fig:stable}(d), $A(t)$ starts from a value
above $C_r$ and terminates at a value below $C_r$, with the upper envelope of
$A(t)$ crossing $C_r$. In Fig.~\ref{fig:stable}(e), $A(t)$ starts from a value
below $C_r$ and terminates at a value below $C_r$. In Fig.~\ref{fig:stable}(f),
$A(t)$ starts from a value above $C_r$ and terminates at a value above $C_r$.
In Figs.~\ref{fig:stable}(c) and \ref{fig:stable}(d), odd intervals are
generated, while in Figs.~\ref{fig:stable}(e) and \ref{fig:stable}(f),
the intervals are even. Between the cases in the same category
[e.g., (c,d) or (e,f)], there is little difference in the statistical
distribution of $T_0$, especially in the long time limit.

We have seen that the equilibrium position $A_{0}$ plays an important role
in terminating the regular oscillations, which can be calculated as
$A_{0}=\langle A(t)\rangle_t$, where $\langle \cdot\rangle_t$ denotes the
average over time. From Fig.~\ref{fig:stable}(b) where the time interval
$T_0$ is displayed as a function of the quantity $A_0-C_r$, we see that the
values of $A_0$ closer to the capacity $C_r$ lead to regular oscillations
with larger values of $T_0$. The most probable values of the distributions
of the even (squares) and odd (stars) $T_0$ values are also indicated in
Fig.~\ref{fig:stable}(b).

\subsection{Mean field theory}

\begin{figure}[h!]
\centering
\includegraphics[width=\linewidth]{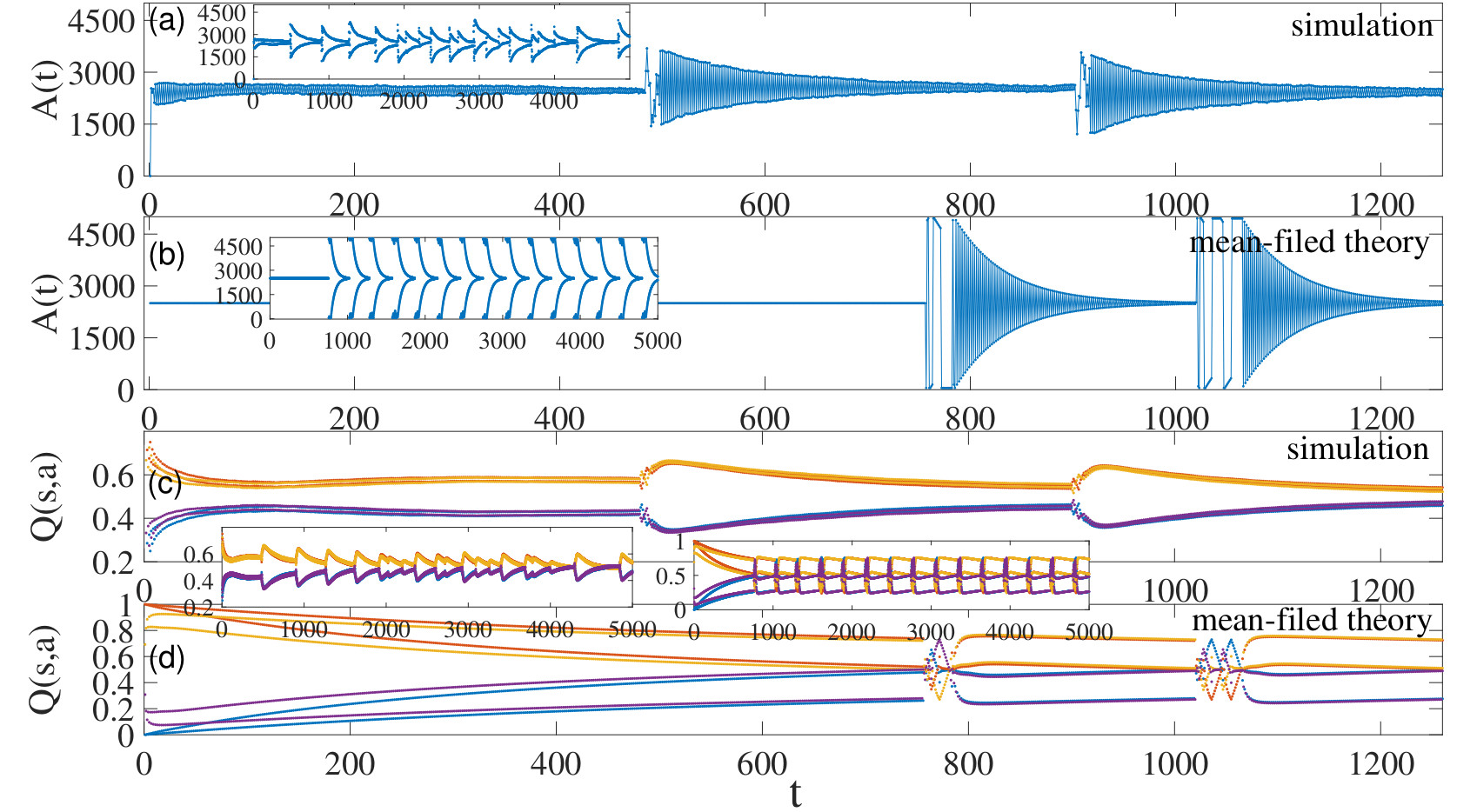}
\caption{ \textbf{Comparison of dynamical evolution of the system obtained
from simulation and mean-field theory.} The attendance $A(t)$ obtained from
(a) multiagent simulation, and (b) numerical solution of
Eqs.~(\ref{eq:Qsas}-\ref{eq:10}). (c,d) The corresponding results of the
elements of $\mathbf{Q}$ from multi-agent simulation and from numerical
solution, respectively. The insets in (a-d) show the corresponding time
series of $A(t)$ and $\mathbf{Q}$ in a large time regime. The parameters
are $\alpha=0.9$, $\gamma=0.9$, $\epsilon=0.02$, and $N=5001$.}
\label{fig:numAt}
\end{figure}

We develop a mean-field analysis to capture the main features of the dynamical
evolution of the multi-agent AI minority system. We assume that the agents
empowered with reinforced learning are identical and share the same matrix
$\mathbf{Q}$. The dynamical evolution of $A(t)$ can be described by the
following equation:
\begin{equation}\label{eq:MF}
\frac{dA(t)}{dt}=\epsilon\frac{N}{2}+(1-\epsilon)[A(t)\Theta(X_1)
+(N-A(t))\Theta(X_2)]-A(t),
\end{equation}
where the first item $\epsilon N/2$ is the number of agents that act randomly
with probability $\epsilon$, half of which select $+1$. The second item
indicates the number of agents that act based on the matrix $\mathbf{Q}$ with
probability $1-\epsilon$, which include agents that stay in the $+1$ state and
those that transition from $-1$ to $+1$. $\Theta(X)$ denotes the step function:
$\Theta(X)=0$ for $X<0$, $\Theta(X)=1/2$ for $X=0$, and $\Theta(X)=1$ for $X>0$.
The quantities $X_1$ and $X_1$ are defined as
$X_1\equiv Q_t(s_+,a_+)-Q_t(s_+,a_-)$, and
$X_2\equiv Q_t(s_-,a_+)-Q_t(s_-,a_-)$.

The elements of the matrix $\mathbf{Q}$ are updated according to the following
rules:
\begin{eqnarray}
\frac{dQ_t(s_+,a_+)}{dt}&=&\alpha[R\Theta(X_3)+\gamma Q_t^{\mathrm{max}}-Q_t(s_+,a_+)][(1-\epsilon)\Theta(X_1)+\frac{1}{2}\epsilon],\label{eq:7}\\
\frac{dQ_t(s_+,a_-)}{dt}&=&\alpha[R\Theta(-X_3)+\gamma Q_t^{\mathrm{max}}-Q_t(s_+,a_-)][(1-\epsilon)\Theta(-X_1)+\frac{1}{2}\epsilon],\label{eq:8}\\
\frac{dQ_t(s_-,a_+)}{dt}&=&\alpha[R\Theta(X_3)+\gamma Q_t^{\mathrm{max}}-Q_t(s_-,a_+)][(1-\epsilon)\Theta(X_2)+\frac{1}{2}\epsilon],\label{eq:9}\\
\frac{dQ_t(s_-,a_-)}{dt}&=&\alpha[R\Theta(-X_3)+\gamma Q_t^{\mathrm{max}}-Q_t(s_-,a_-)][(1-\epsilon)\Theta(-X_2)+\frac{1}{2}\epsilon],\label{eq:10}
\end{eqnarray}
where, $X_3\equiv N-2A(t)$, the step function $\Theta(X_3)$ indicates whether
or not the agents gain a reward, $Q_t^{\mathrm{max}}$ is the expected value
after action. Specifically, we have
$Q_t^{\mathrm{max}}=\max[Q_t(s_+,a_+),Q_t(s_+,a_-)]$ in Eqs.~(\ref{eq:7})
and (\ref{eq:9}) for the agents who take action to transition to $+1$.
Similarly, $Q_t^{\mathrm{max}}=\max[Q_t(s_-,a_+),Q_t(s_-,a_-)]$ in
Eqs.~(\ref{eq:8}) and (\ref{eq:10}) is for agents taking action to
transition to the state $-1$.

The dynamical evolution of the system can thus be assessed either through
simulation, as presented in Figs.~\ref{fig:numAt}(a) and \ref{fig:numAt}(c),
or through the mean-field equations Eqs.~(\ref{eq:Qsas}-\ref{eq:10}), as
shown in Figs.~\ref{fig:numAt}(b) and \ref{fig:numAt}(d). A comparison
between These results indicates that the mean-field equations
Eqs.~(\ref{eq:Qsas}-\ref{eq:10}) capture the main features of the collective
dynamics of the AI minority system, which are regular oscillations with
converging amplitude and intermittent bursts of failure.

\section{Discussion} \label{sec:discussion}

Complex resource allocation systems with a large number of interacting
components are ubiquitous in the modern society. Optimal performance of
such a system is typically measured by uniform and even utilization of all
available resources by the users. Often this is not possible due to the
phenomenon of herding that can emerge spontaneously in the evolution
of the system, in which most agents utilize only a few resources, leaving the
vast majority of the remaining resources little exploited~\cite{PBC:2000,
Vazquez:2000,GL:2002,ZWZYL:2005,EZ:2000,LK:2004,WYZJZW:2005,ZYZXLW:2005,
HWGW:2006,HZDHL:2012,ZHDHL:2013,DHHL:2014}. The heading behavior can
propagate through the system, as the few heavily used resources would
be depleted quickly, directing most agents to another possibly small set
of resources, which would be depleted as well, and so on. A final outcome
is the total collapse of the entire system. An important
goal in managing a complex resource allocation system is to
devise effective strategies to prevent the herding behavior from
occurring. We note that similar behaviors occur in
economics~\cite{BA:1992,CB:2000,AK:2012,MS:2008}. Thus any effective
methods to achieve optimal performance of resource allocation systems
can potentially be generalized to a broader context.

Mathematically, a paradigm to describe and study the dynamics of complex
resource allocation is minority games, in which a large number agents
are driven to seek the less used resources based on available information
to maximize payoff. In the minority game framework, a recent work
addressed the problem of controlling heading~\cite{ZHWSL:2016} using the
pinning method that had been studied in controlling collective dynamics
such as synchronization in complex networks~\cite{WC:2002,LWC:2004,CLL:2007,
XLCCY:2007,TWF:2009,PF:2009,YCL:2009,ZHDHL:2013}, where the dynamics of a
small number of nodes are ``pinned'' to some desired behavior. In developing
a pinning control scheme, the fraction of agents chosen to hold a fixed
state and the structure of the pinned agents are key issues. For the
minority game system, during the time evolution, fluctuations that contain
characteristically distinct components can arise: intrinsic and systematic,
and this allows one to design a control method based on separated control
variables~\cite{ZHWSL:2016}. A finding was that biased pinning
control pattern can lead to an optimal pinning fraction that minimizes the
system fluctuations, and this holds regardless of the network topologies.

Any control based method aiming to suppress or eliminate herding requires
external input. The question we address in this paper is whether it would
be possible to design a ``smart'' type of resource allocation systems that
can sense the potential emergence of herding and adjust the game strategy
accordingly to achieve the same goal but without any external intervention.
Our answer is affirmative. In particular, we introduce AI into
the minority game system in which the agents are ``intelligent'' and
empowered with reinforced learning. Exploiting a popular learning algorithm
in AI, Q-learning, we find that the collective dynamics can evolve to the
optimal state in a self-organized fashion, which is effectively immune
from any herding behavior. Due to the complex dynamics, the evolution toward
the optimal state is not uninterrupted: there can be intermittent bursts of
failures. However, because of the power of self-learning, once a failure event
has occurred, the system can self-repair or self-adjust to start a new process
of evolution toward the optimal state, free of herding. A finding is that two
distinct types of the probability distribution of the intervals of free
evolution (the time interval between two successive failure events) arise,
depending on the parity of the system state. We provide a physical analysis
and derive mean-field equations to understand these behaviors. AI has become
increasingly important and has been universally applied to all aspects of
the modern society. Our work demonstrates, for the first time, that the
marriage of AI with complex systems can generate optimal performance without
the need of external control or intervention.

\section*{Acknowledgements}

We thank Dr.~J.-Q. Zhang and Prof.~Z.-X. Wu for helpful discussions.
This work was supported by the NSF of China under Grants No.~11275003, No.~11775101,
and the Fundamental Research Funds for the Central Universities under Grant
No.~lzujbky-2016-123. YCL would like to acknowledge support from the Vannevar
Bush Faculty Fellowship program sponsored by the Basic Research Office of
the Assistant Secretary of Defense for Research and Engineering and funded
by the Office of Naval Research through Grant No.~N00014-16-1-2828.

\newpage

\section*{Appendix}

\subsection*{Convergence mechanism of $A(t)$}

Typically, after a failure burst, $A(t)$ will converge to the value
corresponding to the optimal system state. The mechanism of convergence can be
understood, as follows. The essential dynamical event responsible
for the convergence is the change of agents from being speculative to being
self-satisfied within the training time. If the inequalities
$Q(s_+,a_+)<Q(s_+,a_-)$ and $Q(s_-,a_-)<Q(s_-,a_+)$ hold, the agent is
speculative and wins the game all the time as a result of the state transition.
Otherwise, for $Q(s_+,a_+)>Q(s_+,a_-)$ and $Q(s_-,a_-)>Q(s_-,a_+)$, the agent
is self-satisfied and wins and loses the game alternatively.

Consider a speculative agent. Assume that its state is $r=+1$ at the
current time step. The agent selects $r=+1$ with the probability
$\epsilon/2$ and updates $Q(s_+,a_+)$ with reward or selects $r=-1$
with the probability $\epsilon/2+(1-\epsilon)$ and updates
$Q(s_+,a_-)$ without reward. If the agent selects $r=+1$, the game will be
lost, but the value of $Q(s_+,a_+)$ can increase. At the next time step,
the agent selects $r=-1$ and loses the game, and $Q(s_+,a_-)$ will decrease.
As a result, the inequality $Q(s_+,a_+)>Q(s_+,a_-)$ holds with the probability
$\epsilon/2$. That is, the probability that a speculative agent
changes to a self-satisfied one is approximately $\epsilon/2$.

Now consider a self-satisfied agent in the $r=+1$ state at the current time
step. The agent selects $r=+1$ with the probability
$\epsilon/2+(1-\epsilon)$ and updates $Q(s_+,a_+)$ with two stable
solutions ($Q_1^*$ and $Q_2^*$), or the agent selects $r=-1$ with the
probability $\epsilon/2$. The agent selects $r=-1$ from the two stable
solutions $Q_1^*$ or $Q_2^*$ with the respective probability $1/2$. If the agent
is associated with the smaller stable solution $Q_2^*$, then $Q(s_+,a_-)$
will decrease. As a result, the agent remains to be self-satisfied. If the
agent is associated with the larger stable solution $Q_1^*$, then $Q(s_+,a_-)$
will increase due to reward, and the inequality $Q(s_+,a_-)>Q(s_+,a_+)$ holds
with the probability $1/2$. At the same time, if $Q(s_-,a_+)>Q(s_-,a_-)$,
the probability is approximately equal to $1/2$, and the self-satisfied agent
successfully becomes a speculative agent. Otherwise, the self-satisfied agent
remains to be self-satisfied. That is, the probability that a self-satisfied
agent changes to being speculative is approximately
$\epsilon/16 \ll \epsilon/2$. As a result, $A(t)$
will converge to $C_r$ asymptotically.

\begin{figure}[h!]
\centering
\includegraphics[width=\linewidth]{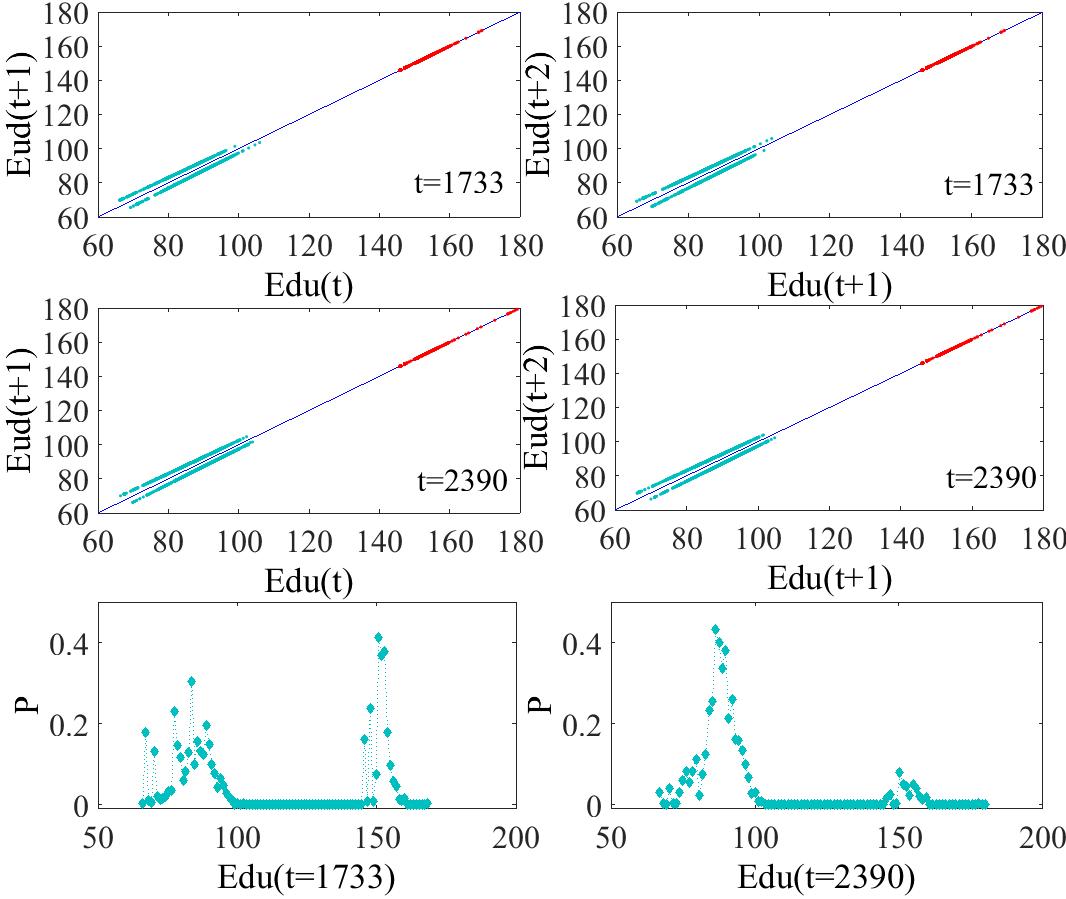}
\caption{ \textbf{Euclidean distance for all agents empowered with reinforced
learning}. The distances are indicated by the two red arrow positions in
Fig.~\ref{fig:sketch}(a). (a-d) Evolution of the Euclidean distance at
three adjacent time steps: $t+1$ vs $t$ and $t+2$ vs $t+1$. The blue line
is $x=y$. Panels (a,b) correspond to the left arrow, and (c,d) to the right
arrow. (e,f) The corresponding distributions of the Euclidean distance.}
\label{fig:Euclidean}
\end{figure}

\paragraph*{Two types of agents in the phase space}
For the AI minority game system, we can construct the phase space, in which
the two types of agents can be distinguished. We define the Euclidean distance
for the $\mathbf{Q}$ matrix of each agent as the square root of the sum of all
the matrix elements. For the two positions indicated by the red arrows in
Fig.~\ref{fig:sketch}(a), Figs.~\ref{fig:Euclidean}(a-d) show the
relationship of Euclidean distance at three adjacent time steps. We see that
the agents can be distinguished and classified into two categories through the
Euclidean distance, where the self-satisfied and the speculative agents
correspond to the top and bottom sides of the line $x=y$ and on the line $x=y$,
respectively. The reason that the speculative agents change their state while
the self-satisfied agents remain in their state lies in the property of the
elements of the $\mathbf{Q}$ matrix. In particular, after the system reaches a
steady state after training, for the speculative agents, the following
inequalities hold: $Q(s_+,a_+)<Q(s_+,a_-)$ and $Q(s_-,a_-)<Q(s_-,a_+)$, while
for the self-satisfied agents, the inequalities are $Q(s_+,a_+)>Q(s_+,a_-)$ and
$Q(s_-,a_-)>Q(s_-,a_+)$. Since the values of the matrix elements $Q(s_+,a_+)$
and $Q(s_-,a_-)$ associated with the self-satisfied agents are between $Q_1^*$
and $Q_2^*$, the Euclidean distance of these agents rolls over on the line
$x=y$ at the adjacent time. However, the elements $Q(s_+,a_-)$ and $Q(s_-,a_+)$
associated with the speculative agents reach only the stable solution $Q_3^*$.
As a result, the Euclidean distance of these agents remain unchanged.
Figures~\ref{fig:Euclidean}(e) and \ref{fig:Euclidean}(f) show that the
Euclidean distances for the agents display a two-peak distribution,
corresponding to the two types. The peak height on the left hand side
increases with time, while that on the right hand side decreases.


\bibliographystyle{iopart-num}
\bibliography{MGRL}

\providecommand{\newblock}{}
\begin{thebibliography}{10}
\expandafter\ifx\csname url\endcsname\relax
  \def\url#1{{\tt #1}}\fi
\expandafter\ifx\csname urlprefix\endcsname\relax\def\urlprefix{URL }\fi
\providecommand{\eprint}[2][]{\url{#2}}

\bibitem{Kauffman:book}
Kauffman S~A 1993 {\em The origins of order: Self-organization and selection in
  evolution\/} (Oxford university press)

\bibitem{Levin:1998}
Levin S~A 1998 {\em Ecosys.\/} {\bf 1} 431--436

\bibitem{AAP:book}
Arthur W~B, Durlauf S~N and Lane D~A 1997 {\em The economy as an evolving
  complex system II\/} vol~28 (Addison-Wesley Reading, MA)

\bibitem{CZ:1997}
Challet D and Zhang Y~C 1997 {\em arXiv preprint adap-org/9708006\/}

\bibitem{NPS:2000}
Nowak M~A, Page K~M and Sigmund K 2000 {\em Science\/} {\bf 289} 1773--1775

\bibitem{RCS:2009}
Roca C~P, Cuesta J~A and S{\'a}nchez A 2009 {\em Phys. Rev. E\/} {\bf 80}
  046106

\bibitem{PD:2012}
Press W~H and Dyson F~J 2012 {\em Proc. Nat. Acad. Sci. (UDA)\/} {\bf 109}
  10409--10413

\bibitem{Arthur:1994}
Arthur W~B 1994 {\em Ame. Econ. Rev.\/} {\bf 84} 406--411

\bibitem{CM:1999}
Challet D and Marsili M 1999 {\em Phys. Rev. E\/} {\bf 60} R6271

\bibitem{SMR:1999}
Savit R, Manuca R and Riolo R 1999 {\em Phys. Rev. Lett.\/} {\bf 82} 2203

\bibitem{PBC:2000}
Paczuski M, Bassler K~E and Corral {\'A} 2000 {\em Phys. Rev. Lett.\/} {\bf 84}
  3185

\bibitem{CMM:2008}
Challet D, Martino A~D and Marsili M 2008 {\em J. Stat. Mech. Theo. E.\/}
  L04004

\bibitem{Moro:2004}
Moro E 2004 {\em Advances in Condensed Matter and Statistical Physics\/} (Nova
  Science Publishers) chap The Minority Games: An Introductory Guide

\bibitem{CMZ:2005}
Challet D, Marsili M and Zhang Y~C 2005 {\em Minority Games\/} Oxford Finance
  (Oxford University Press)

\bibitem{YZ:2008}
Yeung C~H and Zhang Y~C 2009 Minority games {\em Encyclopedia of Complexity and
  Systems Science\/} ed Meyers R~A (Springer New York) pp 5588--5604

\bibitem{ZWZYL:2005}
Zhou T, Wang B~H, Zhou P~L, Yang C~X and Liu J 2005 {\em Phys. Rev. E\/} {\bf
  72} 046139

\bibitem{EZ:2000}
Eguiluz V~M and Zimmermann M~G 2000 {\em Phys. Rev. Lett.\/} {\bf 85} 5659

\bibitem{TCHJ:2005}
Lo T~S, Chan K~P, Hui P~M and Johnson N~F 2005 {\em Phys. Rev. E\/} {\bf 71}
  050101

\bibitem{JHH:1999}
Johnson N~F, Hart M and Hui P~M 1999 {\em Physica A\/} {\bf 269} 1--8

\bibitem{HJJH:2001}
Hart M, Jefferies P, Johnson N~F and Hui P~M 2001 {\em Physica A\/} {\bf 298}
  537--544

\bibitem{Marsili:2001}
Marsili M 2001 {\em Physica A\/} {\bf 299} 93 -- 103 ISSN 0378-4371 application
  of Physics in Economic Modelling

\bibitem{BMFM:2008}
Bianconi G, Martino A~D, Ferreira F~F and Marsili M 2008 {\em Quant. Financ.\/}
  {\bf 8} 225--231

\bibitem{XWHZ:2005}
Xie Y~B, B~H~Wang C~K~H and Zhou T 2005 {\em Eur. Phys. J. B\/} {\bf 47} 587

\bibitem{ZZZH:2005}
Zhong L~X, Zheng D~F, Zheng B and Hui P~M 2005 {\em Phys. Rev. E\/} {\bf 72}(2)
  026134

\bibitem{ATBK:2004}
Anghel M, Toroczkai Z, Bassler K~E and Korniss G 2004 {\em Phys. Rev. Lett.\/}
  {\bf 92}(5) 058701

\bibitem{LCHJ:2004}
Lo T~S, Chan H~Y, Hui P~M and Johnson N~F 2004 {\em Phys. Rev. E\/} {\bf 70}(5)
  056102

\bibitem{Slanina:2001}
Slanina F 2001 {\em Physica A\/} {\bf 299} 334

\bibitem{KSB:2000}
Kalinowski T, Schulz H~J and Birese M 2000 {\em Physica A\/} {\bf 277} 502

\bibitem{MMM:2004}
Martino A~D, Marsili M and Mulet R 2004 {\em Europhys. Lett.\/} {\bf 65} 283

\bibitem{BMM:2007}
Borghesi C, Marsili M and Miccich\`e S 2007 {\em Phys. Rev. E\/} {\bf 76}(2)
  026104

\bibitem{AK:2002}
Galstyan A and Lerman K 2002 {\em Phys. Rev. E\/} {\bf 66}(1) 015103

\bibitem{ZHDHL:2013}
Zhang J~Q, Huang Z~G, Dong J~Q, Huang L and Lai Y~C 2013 {\em Phys. Rev. E\/}
  {\bf 87}(5) 052808

\bibitem{ZHWSL:2016}
Zhang J~Q, Huang Z~G, Wu Z~X, Su R~Q and Lai Y~C 2016 {\em Sci. Rep.\/} {\bf 6}

\bibitem{HZDHL:2012}
Huang Z~G, Zhang J~Q, Dong J~Q, Huang L and Lai Y~C 2012 {\em Sci. Rep.\/} {\bf
  2} 703

\bibitem{DHHL:2014}
Dong J~Q, Huang Z~G, Huang L and Lai Y~C 2014 {\em Phys. Rev. E\/} {\bf 90}
  062917

\bibitem{AA:book}
Richard S~Sutton A~G~B 1998 {\em Reinforcement Learning: An Introduction\/}
  vol~21 (Cambridge MA: The MIT press)

\bibitem{Bellman:1957}
Bellman R~E 1957 {\em Dynamic Programing\/} (Princeton, New Jersey: Princeton
  University Press)

\bibitem{Sutton:1998}
Sutton R~S 1998 {\em Mach. Learning\/} {\bf 3} 9--44

\bibitem{Watkins:1989}
Watkins C~J~C 1989 {\em PhD Thesis Cambridge University\/}

\bibitem{WD:1992}
Watkins C~J~C~H and Dayan P 1992 {\em Mach. Learning\/} {\bf 8} 279--292

\bibitem{AA:2012}
Kianercy A and Galstyan A 2012 {\em Phys. Rev. E\/} {\bf 85}(4) 041145

\bibitem{PA:2003}
Potapov A and Ali M~K 2003 {\em Phys. Rev. E\/} {\bf 67}(2) 026706

\bibitem{SC:2003}
Sato Y and Crutchfield J~P 2003 {\em Phys. Rev. E\/} {\bf 67}(1) 015206

\bibitem{AA:2013}
Kianercy A and Galstyan A 2013 {\em Phys. Rev. E\/} {\bf 88}(1) 012815

\bibitem{DW:2016}
Das R and Wales D~J 2016 {\em Phys. Rev. E\/} {\bf 93}(6) 063310

\bibitem{Vazquez:2000}
V{\'a}zquez A 2000 {\em Phys. Rev. E\/} {\bf 62} R4497

\bibitem{GL:2002}
Galstyan A and Lerman K 2002 {\em Phys. Rev. E\/} {\bf 66} 015103

\bibitem{LK:2004}
Lee S and Kim Y 2004 {\em J. Korean. Phys. Soc.\/} {\bf 44} 672--676

\bibitem{WYZJZW:2005}
Wang J, Yang C~X, Zhou P~L, Jin Y~D, Zhou T and Wang B~H 2005 {\em Physica A\/}
  {\bf 354} 505--517

\bibitem{ZYZXLW:2005}
Zhou P~L, Yang C~X, Zhou T, Xu M, Liu J and Wang B~H 2005 {\em New Math. Nat.
  Comp.\/} {\bf 1} 275--283

\bibitem{HWGW:2006}
Huang Z~G, Wu Z~X, Guan J~Y and Wang Y~H 2006 {\em Chin. Phys. Lett.\/} {\bf
  23} 3119

\bibitem{BA:1992}
Banerjee A~V 1992 {\em Q. J. Econ.\/}  797--817

\bibitem{CB:2000}
Cont R and Bouchaud J~P 2000 {\em Macroecon. Dyn.\/} {\bf 4} 170--196

\bibitem{AK:2012}
Ali S~N and Kartik N 2012 {\em Econ. Theor.\/} {\bf 51} 601--626

\bibitem{MS:2008}
Morone A and Samanidou E 2008 {\em J. Evol. Econ\/} {\bf 18} 639--646

\bibitem{WC:2002}
Wang X~F and Chen G 2002 {\em Physica A\/} {\bf 310} 521--531

\bibitem{LWC:2004}
Li X, Wang X and Chen G 2004 {\em IEEE Trans. Circ. Sys.\/} {\bf 51} 2074--2087

\bibitem{CLL:2007}
Chen T, Liu X and Lu W 2007 {\em IEEE Trans. Circ. Sys.\/} {\bf 54} 1317--1326

\bibitem{XLCCY:2007}
Xiang L, Liu Z, Chen Z, Chen F and Yuan Z 2007 {\em Physica A\/} {\bf 379}
  298--306

\bibitem{TWF:2009}
Tang Y, Wang Z and Fang J~a 2009 {\em Chaos\/} {\bf 19} 013112

\bibitem{PF:2009}
Porfiri M and Fiorilli F 2009 {\em Chaos\/} {\bf 19} 013122

\bibitem{YCL:2009}
Yu W, Chen G and L{\"u} J 2009 {\em Automatica\/} {\bf 45} 429--435

\end{thebibliography}

\end{document}